\documentstyle[prl,epsfig,aps,multicol]{revtex}
\begin{document}
\draft
\title{\bf Dynamical Criterion for Quantum Chaos : Entropy Production in Subsystems}
\author{Avijit Lahiri\thanks{lahiri@cal2.vsnl.net.in}}
\address{\bf Dept of Physics, Vidyasagar Evening College, Kolkata 700 006, INDIA}
\maketitle
\begin{abstract}
We examine the conjecture that  entropy production in subsystems of a given system can be used as a dynamical criterion for quantum chaos in the latter. Numerical results are presented for finite dimensional spin systems as also for the quantum baker's map. Of especial importance is the power spectrum of the entropy production which gets progressively more and more broad-banded as the degree of correlation in the Hamiltonian matrix is made to decrease.
\end{abstract}
\begin {multicols}{2}
More than twenty years after the discovery of dynamical localisation \cite{qloc}, possible signatures of quantum chaos are still under active exploration. While investigations have lent overwhelming support to a set of sharp conjectures and predictions relating to stationary aspects of the problem like level statistics and eigenvector distributions (see, e.g., \cite{Hake,Stockman,Casati-coll}), corresponding {\it dynamical} criteria of quantum chaos are only recently beginning to take shape. One recalls that a simple-minded search for chaotic features in quantum time evolution of a system along lines similar those in classical evolution should yield negative results in consequence of the linearity of the Schr\"odinger equation. A more fruitful approach, on the other hand, would be to focus on {\it reduced states} of subsystems. Recent investigations have focussed on the reduced entropy of subsystems~\cite{Arul1} as an important indicator in the context of quantum chaos. For a set of relevant results on reduced entropy as a measure of correlation in subsystems, see~\cite{entropy1,entropy2}
 
The stationary criteria of level statistics and eigenvector distribution share the satisfactory feature that they are specifically quantal in nature, having no direct reference to the semiclassical limit. Dynamical criteria like the spreading of wavepackets  examined in earlier investigations (see, e.g.,~\cite{Heller,Fishman}) on quantum chaos are not entirely satisfactory on this count since they refer to time scales pertaining to the semiclassical limit. By contrast, the dynamic criterion of entropy production in subsystems retains its relevance in the deep quantum regime as well as in the semiclassical one.

Quantum chaos has acquired a special relevance in the context of quantum information processing and quantum computation \cite{qcomp} in that it reflects in the manner in which {\it decoherence} and {\it entanglement}\cite {entang1,entang2,entang3} arise in a system of qbits. Entropy production can be used in this context as a natural measure of decoherence, and it seems appropriate to look at various features of entropy production in order to formulate a sufficiently satisfactory and universal dynamical criterion for quantum chaos. The degree of entanglement, which is  represented by entropy production under certain circumstances, has also been shown to be sensitive to the presence of classical chaos\cite{entropy1,Arul1,Arul2}. 
The present paper folows this line of approach and aims to point out that  features of entropy production are relevant in characterising quantum chaos, one such useful feature being  the time dependence of entropy production as revealed by its {\it{power spectrum}}. 

More precisely, starting from an initial state $\rho(0)$ of a system $S$ made up of subsystems $S_1$ and $S_2$, we let the system evolve for a time $t$ and obtain the reduced state ${\rho}_R(t)$ of $S_1$ on taking partial trace over $S_2$. We then look at the time dependence of ${\rho}_R$. Even when one starts from a pure state $\rho(0)$, ${\rho}_R$ turns out more often than not to correspond to a mixed state and there takes place, in general, a decoherence with the passage of time. As mentioned above, one can measure the extent of decoherence by the reduced von Neumann entropy
\begin{equation}
s_R(t)=-trace({\rho}_R(t)ln{\rho}_R(t)).\label{entropy}
\end{equation}
We compare features of time dependence of $s_R(t)$ for systems conforming to the predictions of random matrix theory (RMT) with the corresponding features for a `regular' (see below) syestm. We present evidence that while the former is characterised by a broad-band power spectrum, the latter involves sparsely distributed peaks, indicative of a quasiperiodic time dependence.The system we consider is the simple one of an interacting assembly ($S$) of $n$ spins or, in other words, an $n$-qbit system. The state space has dimension $N=2^n$, and the Hamiltonian determining the time evolution of $S$ is an $N\times N$ matrix. While quantum chaos in the true sense is expected to emerge only in the limit $N\rightarrow\infty$, we consider finite (and rather low) values of $N$ here so as to present in as simple a context as possible the features we aim to underline in this note. Finite dimensional considerations are, moreover relevant in numerous contexts relating to quantum computation.

Thus, we consider, first, an $N\times N$ real symmetric Hamiltonian matrix $H_c$ with randomly distributed matrix elements for which the stationary features predicted by RMT are conformed to and, next, a second Hamiltonian matrix $H_r$ that is equivalent to the Harper Hamiltonian on a torus. In the qbit representation the Hamiltonians correspond to strong and long-range interactions among the qbits. Alternatively, either of these can also be interpreted to represent the Hamiltonian of a chain of $N$ fermions 
of the form
 \begin{equation}
\sum a_{ij}({c_i}^{+}c_j+c_i{c_j}^{+}),\label{aij}
\end{equation}
where $a_{ij}=a_{ji}$, and $c_i$, ${c_i}^{+}$ denote fermion annihilation and creation  operators. For $H_c$ , the coefficients $a_{ij}$ are chosen as randomly distributed matrix elements (subject to reality ad hermiticity) while, for $H_r$, we take
\begin{equation}H_r={\gamma}_1T+{\gamma}_2V.\label{H_r}
\end{equation}
Here ${\gamma}_1$ , ${\gamma}_2$ are real constants,
\begin{equation}T_{i,i+1}=T_{i+1,i}=T_{1N}=T_{N1}=1/2~~(i=1,...N-1),\label{Ti_i+1}
\end{equation}and 
\begin{equation} V_{ij}=\cos(2{\pi}j/N){\delta}_{ij}.\label{V_ij}
\end{equation}
This represents a Harper's Hamiltonian
\begin{equation}H_r = {\gamma}_1\cos(2{\pi}p/P) +{\gamma}_2\cos(2{\pi}q/Q),\label{Harper}
\end{equation}
on a torus of area $PQ$ with $N$ basic states ($\hbar=PQ/2{\pi}N$) and with torus boundary conditions ${\omega}_1={\omega}_2=0$. The parameters ${\gamma}_1, {\gamma}_2$ have been set at (resp.) $0.5, 2.5$ in our computations. In the fermion-chain interpretation, $H_r$ represents a Hamiltonian in the $1$-fermion subspace for the chain\cite{Arul2}. However, in the following, we refer to $H_c$ and $H_r$ simply as Hamiltonians of an interacting spin system. 

With the Hamiltonian of the system $S$ defined as above, we focus on a subsystem $S_1$ made up of, say, $p(<n)$ specified spins and look at the reduced density matrx (RDM) given by
\begin{equation}{\rho_R}(t)={Tr}^{(n-p)}[e^{-iHt\over\hbar}\rho(0)e^{iHt\over\hbar}],\label{evolution}
\end{equation}
where $H=H_c$ or $H_r$ as the case may be, and where ${Tr}^{(n-p)}$ indicates partial trace with respect to states of the remaining $(n-p)$ number of spins. We then calculate $s_R(t)$ of Eq.(1). We present below numerical results for $n=8$ ($N=256$), and $p=5$, i.e., a subsystem made up of five spins. For convenience of presentation, we represent all the quantities as dimensionless; in other words, we use arbitrary but consistent units. 
Fig. 1 gives a comparison of the typical time-dependence of $s_R(t)$ for (a) $H_c$ and for (b) $H_r$  with the same value of $\hbar$. For the sake of comparison we subtract a constant term from each matrix element of $H_c$, $H_r$ so as to reduce to zero the mean of all the matrix elements.
\begin{figure*}
\begin{center}
\epsfig{figure=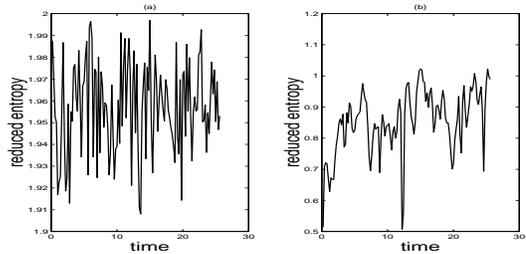,height=3.5cm, width=7cm}
\end{center}
\caption{Entropy production as a function of time for (a) $H_c$, and (b) $H_r$, ($\hbar=.1$ in both cases); note the erratic character of the variation in (a) as opposed to the more regular variation in (b).}
\end{figure*}
Referring to Fig. 1 (see also Fig. 6)  we note the following distinctive features of $s_R(t)$ : (i) there is an initial `transient' regime in which $s_R(t)$ increases more rapidly for $H_c$ (almost instantaneous in the scale of Fig. 1, but discernible in Fig. 6(a) for the quantum baker's map) than for $H_r$; (ii) after the initial transient regime there arises a `steady' state in which $s_R(t)$ fluctuates according to some invariant pattern; however, the mean value around which the fluctuation takes place is, in general, larger for $H_c$ than for $H_r$; (iii) the fluctuation is more erratic for $H_c$ than for $H_r$ - while the latter appears to be quasiperiodic in nature, the former is random by comparison. 
\begin{figure*}
\epsfig{figure=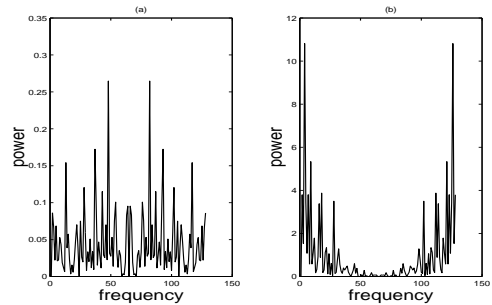,height=4cm, width=6.5cm}
\vskip .5 cm
\caption{Power spectrum of $s_R(t)$ for (a) $H_c$, and (b) $H_r$; the initial transient regime has been ignored while obtaining the power spectrum; note the broad-band nature of the spectrum in (a) in contrast to (b).}
\end{figure*}
The contrast between the modes of fluctuation in the two situations under study is seen more clearly in Fig. 2 where we present the power spectrum of the fluctuations in $s_R$. One finds that the power spectrum for $H_c$ is a broad-band one while that for $H_r$ has a relatively smaller number of sharp peaks (the power spectra have been obtained with the help of a standard FFT package, and the symmetry about the mid-point of the frequency range has no physical relevance, being an artifact resulting from the program). It may be mentioned that the qualitative features indicated above persist against changes in the value of $\hbar$ and the dimension of the subsystem considered. Details will be presented elsewher.

In order to set in perspective the distinction between $H_r$ and $H_c$ one can compute the nearest neighbour level spacing distribution for the two Hamiltonians. One observes in Fig. 3 that the level spacing distribution for $H_c$ is close to the Wigner distribution implied by RMT, while that for $H_r$ deviates to a large extent from the Wigner distribution and is, in fact, closer to the Poisson distribution followed by systems close to integrable classical systems. In other words, there exists a discernible correlation between stationary features implied by the RMT and dynamic features of entropy production in subsystems. Put differently, entropy production can indeed be taken as a valid dynamical criterion for quantum chaos - a criterion, moreover, that makes no direct reference to classical time evolution. While the latter feature is apparent from the manner in which the criterion has been formulated, one also finds support for this assertion from the fact that features of entropy production mentioned above appear even in situations with higher values of $\hbar$ (not shown in our figures) compared to the value used here.
\begin{figure*}
\begin{center}
\epsfig{figure=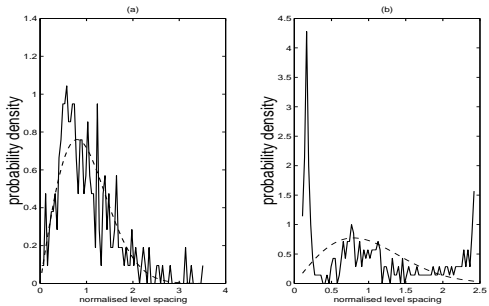,height=4cm, width=6.5cm}
\end{center}
\caption{Nearest neighbour level spacing distribution for (a) $H_c$, and (b) $H_r$; the dotted line compares the distribution in each case with a Wigner distribution with appropriate mean and variance .}
\end{figure*}
A quantum system represented by a Hamiltonian with randomly distributed matrix elements is, in a sense, analogous to classical systems exhibiting `hard' chaos. By contrast, analogs to `regular' classical systems would be ones with Hamiltonian matrices having correlated matrix elements. In between, there exist Hamiltonian matrices with matrix elements having various intermediate degrees of correlation. It would be interesting to see if features of entropy production in subsystems of such systems is, in some sense, intermediate between those of the systems considered above, analogous to the fact that systems characterised by `soft' chaos in the semiclassical limit exhibit level spacing distributions intermediate between the Wigner and Poisson distributions(see e.g.,\cite{Brody}).
 
With this in view we modify our Hamiltonian $H_r$ by replacing a given fraction of matrix elements with randomly distributed ones. To be more precise, we construct a family of Hamiltonian matrices $H(f)$ depending on the parameter $f (0<f<1)$ where increasing values of $f$ correspond to greater degrees of correlation between the matrix elenments. $H(f)$ is obtained from $H_r$ by replacing ${(H_r)}_{mn}$ with randomly distributed numbers for $m,n$ satisfying $|m-n|~>~f N$, where $f$ is some chosen fraction. Plots of $s_R(t)$ for decreasing values of $f$(not shown in our our figures) indicate that decreasing values of the degree of correlation indeed correspond to a gradual transition in features of entropy production from those of regular systems (Fig. 1(b)) to chaotic ones (Fig. 1(a)), being intermediate between the two. This is further corroborated from Figure 4 where we display the corresponding power spectra for two chosen values of $f$. One observes that with decreasing $f$ the power spectrum of entropy production becomes more and more broad-banded.
\begin{figure*}
\begin{center}
\epsfig{figure=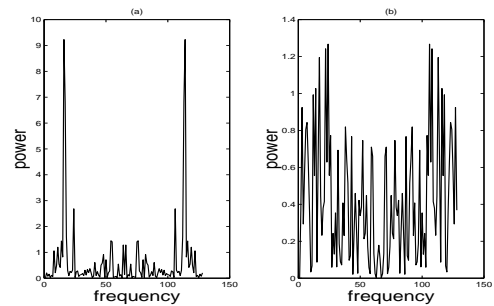,height=4cm, width=6.5cm}
\end{center}
\caption{Power spectrum of $s_R(t)$ with (a) $f=0.9$, and (b) $f=0.8$; in (a), the spectrum becomes more broad-banded with a decrease in correlation among the elements of the Hamiltonian matrix, corresponding to a decrease in $f$.}
\end{figure*}
The fact that the dynamic criterion of chaos indicated in this paper including, in particular, the power spectrum of entropy production, is well correlated with the static criteria implied by RMT is borne out by looking at eigenvector distributions of $H(f)$ for various values of $f$. 

The results are presented in Fig. 5 in which the {\it{residual parameters}}~\cite{residual} for the eigenvectors of $H(f)$ are displayed, once again for a chosen pair of values of $f$. The residual parameter for any given eigenvector is defined as \begin{equation}r={1\over N}(\sum (p_k-{p_k}^{(0)})^2)^{1\over 2},\label{residual}\end{equation}where the set of numbers $p_k$ represent the probability distribution for the elements of the eigenvector and ${p_k}^{(0)}$ stands for a Gaussian distribution with the same mean and variance. Low values of $r$ for all the eigenvectors implies a correspondingly high degree of allegiance to the predictions of the RMT. One observes from the figures that  decreasing values of $f$ indeed corresponds to an increasing degree of conformity to the stationary features implied by RMT.
\begin{figure}
\begin{center}
\epsfig{figure=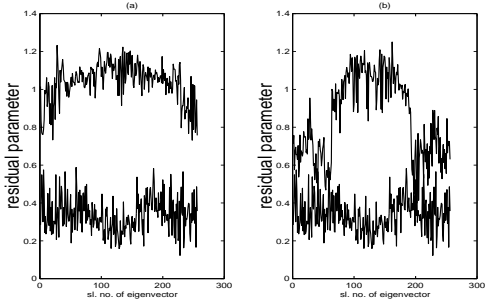,height=4cm, width=6.5cm}
\end{center}
\caption{(a) Residual parameters of eigenvectors for $f=0.9$ (upper trace) compared with those for an uncorrelated Hamiltonian ($f=0$, lower trace); note a large deviation from predictions of RMT; (b) corresponding comparison for $f=0.7$ (upper trace) with the uncorrelated case ($f=0$); the two sets of residual parameters are now closer to each other.}
\end{figure}
The above features of reduced entropy for $H_c$ are also in evidence for the {\it quantum baker's map}~\cite{baker}. Fig. 6 shows (a) the time-dependence of $s_R$, and (b) the power spectrum, which are of the general nature of Fig.s 1(a), 2(a) (as opposed to 1(b), 2(b)) respectively. Interesting details of entropy production for the quantum baker's map, as also the quantum cat map~\cite{cat}, both known to be chaotic in the classical limit, will be presented elsewhere.
\begin{figure}
\begin{center}
\epsfig{figure=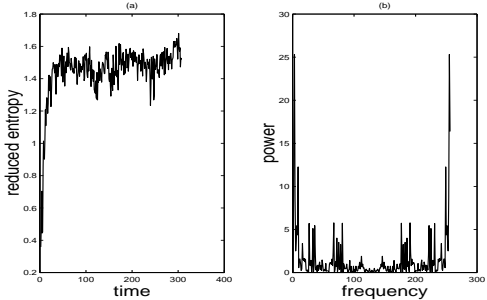,height=4cm, width=6.5cm}
\end{center}
\caption{Entropy production in the quantum baker's map with dimension $N=128$, the subsystem chosen being of dimension $8$; (a) reduced entropy as a function of $t$; note the fast rise in the transient regime; (b) power spectrum, showing an emerging broad-band structure for even the comparatively low value of $N$ chosen .}
\end{figure}
In summary, we have presented evidence that features of entropy production, including the power sprctrum in the steady state, in subsystems of a given system can be used as dynamical criteria of quantum chaos in the latter. Such dynamical criteria complement the stationary criteria discussed in the literature. The features presented show a progressive transition from regular to chaotic systems when the degree of correlation in the underlying Hamiltonian matrix is made to decrease.

While entropy production in subsystems can be loooked upon as an `internal' dynamical criterion for chaos, dynamical features of an `external' nature have also been discussed in the lterature in various related contexts\cite{fidelity,sakagami,Sarben,nag}, namely the distinctive features of chaotic systems in interaction with or perturbed by external ones.   

Other features of entropy production in subsystems, including fractal characteristics, will be presented in a future publication.

\end{multicols}
\end{document}